\documentclass[prl,aps,twocolumn,superscriptaddress]{revtex4-1}

\usepackage{amsmath,amssymb,amsfonts,dsfont,color,graphicx,tabularx,stmaryrd,simplewick,diagbox}
\usepackage[unicode=true,colorlinks=true]{hyperref}
\usepackage{txfonts}
\usepackage[normalem]{ulem}

\hypersetup{linkcolor=blue,citecolor=blue,urlcolor=blue}

\begin{document}

\title{Matrix product states for Hartree-Fock-Bogoliubov wave functions}

\author{Hui-Ke Jin}
\affiliation {Department of Physics TQM, Technische Universit\"{a}t M\"{u}nchen, James-Franck-Straße 1, D-85748 Garching, Germany}

\author{Rong-Yang Sun}
\affiliation{Kavli Institute for Theoretical Sciences, University of Chinese Academy of Sciences, Beijing 100190, China}
\affiliation{Computational Materials Science Research Team, RIKEN Center for Computational Science (R-CCS), Kobe, Hyogo, 650-0047, Japan}
\affiliation{Quantum Computational Science Research Team, RIKEN Center for Quantum Computing (RQC), Wako, Saitama, 351-0198, Japan}

\author{Yi Zhou}
\email{yizhou@iphy.ac.cn}
\affiliation {Beijing National Laboratory for Condensed Matter Physics $\&$ Institute of Physics, Chinese Academy of Sciences, Beijing 100190, China}
\affiliation{Songshan Lake Materials Laboratory, Dongguan, Guangdong 523808, China}
\affiliation{Kavli Institute for Theoretical Sciences, University of Chinese Academy of Sciences, Beijing 100190, China}
\affiliation{CAS Center for Excellence in Topological Quantum Computation, University of Chinese Academy of Sciences, Beijing 100190, China}

\author{Hong-Hao Tu}
\email{hong-hao.tu@tu-dresden.de}
\affiliation{Institut f\"ur Theoretische Physik, Technische Universit\"at Dresden, 01062 Dresden, Germany}

\begin{abstract}
	We provide an efficient and accurate method for converting Hartree-Fock-Bogoliubov wave functions into matrix product states (MPSs). These wave functions, also known as ``Bogoliubov vacua'', exhibit a peculiar entanglement structure that the eigenvectors of the reduced density matrix are also Bogoliubov vacua. We exploit this important feature to obtain their optimal MPS approximation and derive an explicit formula for corresponding MPS matrices. The performance of our method is benchmarked with the Kitaev chain and the Majorana-Hubbard model on the honeycomb lattice. The approach facilitates the applications of Hartree-Fock-Bogoliubov wave functions and is ideally suited for combining with the density-matrix renormalization group method.
\end{abstract}

\maketitle

The pairing of fermions plays a fundamental role in understanding fantastic phenomena in various aspects of physics, such as superconductivity~\cite{Bardeen1957} and superfluidity~\cite{Leggett1975}. The Bardeen-Cooper-Schrieffer (BCS) theory for superconductivity is a celebrated example of this paradigm~\cite{Bardeen1957}. A natural generalization of the BCS theory is the so-called Hartree-Fock-Bogoliubov (HFB) theory~\cite{ring-schuck-book}. It performs variational optimization within a class of states~\cite{Bach1994}, dubbed as quasi-free states or Bogoliubov vacua, which are ground states of fermionic quadratic Hamiltonians. The HFB wave functions, including the Hartree-Fock wave functions (Slater determinants) as a subclass, have been widely applied in diverse fields of physics and chemistry. In particular, Anderson's proposal~\cite{Anderson1987} of a Gutzwiller projected BCS state, which is a dressed version of the HFB wave function, has provided invaluable insights into the high-$T_c$ superconductivity.

Matrix product states (MPSs)~\cite{Klumper1991,Fannes1992}, being the underlying variational ansatz~\cite{Ostlund1995,Rommer1997} of the density-matrix renormalization group (DMRG) method~\cite{White1992,White1993}, form another representative family of many-particle wave functions with broad applications in physics and quantum chemistry~\cite{Verstraete2008,Cirac2009,Schollwock2011,Chan2011,Stoudenmire2012,Orus2014,Wouters2014,Szalay2015}. In this regard, it is natural to ask whether there is a way to connect these two classes of wave functions with each other, especially how to represent one in terms of the other. Recently, several methods for converting HFB wave functions (and their dressed versions) into MPSs have been proposed~\cite{Fishman2015,Tu2020,Jin2020,Petrica2021,Jones2021a,Jones2021b}, enabling us to use them as initial input in DMRG calculations. This approach has the advantage that physically motivated HFB wave functions can guide DMRG to circumvent local minima and accelerate the search of true ground states~\cite{Jin2021a,Aghaei2020,Chen2021,Jin2021b}. While these methods are successful in compressing Hartree-Fock wave functions into MPSs~\cite{Fishman2015,Tu2020,Petrica2021}, their applications to HFB wave functions with fermion pairing are much less satisfactory: For instance, the performance of the previously proposed MPO-MPS method~\cite{Jin2020} relies on the existence of a well localized basis for Bogoliubov quasiparticles, which cannot always be guaranteed. Another method proposed in Ref.~\cite{Petrica2021} generally needs an extra copy of the HFB state to obtain a Slater determinant as advance-preparation. This will double the entanglement of the original HFB state, requiring a much larger bond dimension for the MPS and significantly affecting the accuracy in large-scale computations.

In this Letter, we propose an efficient and accurate MPS representation for HFB wave functions (hereafter referred to as the Pfaffian method). Explicitly, we utilize the correlation matrix technique to diagonalize the reduced density matrix of HFB wave functions. The diagonalized reduced density matrix identifies the degrees of freedom associated with most significant weight in the bipartite entanglement entropy, and leads to a natural MPS decomposition and compression scheme. Furthermore, we reveal that the eigenvectors of the reduced density matrix are all Bogoliubov vacua, as long as a suitable canonical basis is chosen. This peculiar property allows us to derive a Pfaffian formula for the MPS matrices. Similar to the method for Slater determinants~\cite{Petrica2021}, the Pfaffian method is also easily parallelizable. The benchmark results on the Kitaev chain and the Majorana-Hubbard model on the honeycomb lattice suggest advantage over previous methods and provide a promising prospect for combining HFB and DMRG methods.

{\em Compressing Bogoliubov vacua into MPS} --- The system of our concern consists of a quadratic Hamiltonian of fermions, whose creation and annihilation operators are $a^\dagger_j$ and $a_j$, $j=1,\ldots,N$, with $N$ being the number of modes. In the presence (absence) of pairing terms, the ground state $|\Psi\rangle$ is a HFB (Hartree-Fock) wave function. Such a state belongs to the so-called fermionic Gaussian states~\cite{Bravyi2005}, which are fully characterized by the $2N\times 2N$ correlation matrix
\begin{equation}
	\Omega =
	\begin{pmatrix}
		\langle a_{j}^{\dagger }a_{i}\rangle & \langle a_{j}a_{i}\rangle \\
		\langle a_{j}^{\dagger }a_{i}^{\dagger }\rangle & \langle
		a_{j}a_{i}^{\dagger }\rangle
	\end{pmatrix}_{1\leq i,j \leq N},
	\label{eq:ComplexCorrMat}
\end{equation}
where the expectation value is taken with respect to the ground state $|\Psi\rangle$. Thanks to Wick’s theorem, higher-order correlators are completely determined by $\Omega$.

In order to find an optimal MPS representation, the reduced density matrix of $|\Psi\rangle$ is exploited as well as its eigenvalues and eigenvectors (Schmidt vectors). Consider a subsystem $A$ composing of $M$($<N$) fermionic modes with index $j=1,\ldots,M$, the corresponding reduced density matrix $\rho_A$ is a mixed fermionic Gaussian state. The $2M\times 2M$ correlation matrix $\Omega_A$ characterizing $\rho_A$ takes the same form as Eq.~(\ref{eq:ComplexCorrMat}), except that the indices are restricted to $1\leq i,j\leq M$~\cite{Peschel2003}. The diagonalization of $\Omega_A$ can be achieved via a Bogoliubov transformation within subsystem $A$,
\begin{equation}
	\begin{pmatrix}
		d^{\dagger}_A & d_A
	\end{pmatrix}
	=
	\begin{pmatrix}
		a^{\dagger} & a
	\end{pmatrix}
	\begin{pmatrix}
		U_A & V^{*}_A \\
		V_A & U^{*}_A
	\end{pmatrix},
	\label{eq:Bogoliubov_trans}
\end{equation}
where (and hereafter) the creation and annihilation operators without indices represent row vectors, e.g., $d_A = (d_{A,1}, \ldots, d_{A,M})$. The Bogoliubov matrix obeys the relations $U_{A}^\dag U_{A} + V_{A}^\dag V_{A} = \mathds{1}$ and $U_{A}^\dag V^{*}_{A} + V_{A}^\dag U^{*}_{A} = \mathbf{0}$.
The eigenvalues of $\Omega_A$ come in pairs and take the form of $\{\Lambda_{A,p},1-\Lambda_{A,p}\}$, where $p=1,\ldots,M$ and $\Lambda_{A,p}\in[0,1]$. The corresponding eigenvectors are the $p$-th and ($M+p$)-th columns of the Bogoliubov matrix in Eq.~\eqref{eq:Bogoliubov_trans}, respectively. Interchanging the $p$-th and ($M+p$)-th columns in the Bogoliubov matrix in Eq.~\eqref{eq:Bogoliubov_trans} corresponds to a particle-hole transformation $d^{\dagger}_{A,p} \leftrightarrow d_{A,p}$.

The Bogoliubov modes bring the reduced density matrix $\rho_A$ into a simple form
\begin{equation}
	\rho_A=\prod_{p=1}^{M}\left[\Lambda_{A,p}{}d^\dag_{A,p}{}d_{A,p}+(1-\Lambda_{A,p})d_{A,p}d^\dagger_{A,p}\right]
	\label{eq:rhoA},
\end{equation}
where $d^\dag_{A,p}{}d_{A,p}$ and $d_{A,p}d^\dag_{A,p}{}$ project a state onto the occupied and the empty states of the $d_{A,p}$ mode, respectively. In this form, $\rho_A$ can be factorized as a direct product of independent two-level systems. The eigenvectors of $\rho_A$ are the Fock basis of $d_{A}$-modes,
\begin{equation}
	|m_A\rangle=(d^\dagger_{A,1})^{m_1}(d^\dagger_{A,2})^{m_2}\cdots(d^\dagger_{A,M})^{m_{M}}|0\rangle_{d_A},
	\label{eq:vacua_d}
\end{equation}
where $|0\rangle_{d_A}$ is the vacuum of $d_A$-modes, i.e., a Bogoliubov vacuum, $m_i=0,1$ is the occupation number of the $i$-th $d_A$-mode, and $m_A = \{m_1,\ldots,m_{M}\}$ labels these Schmidt vectors. It is worth emphasizing that such a Schmidt vector can be transformed to a Bogoliubov vacuum by performing a suitable particle-hole transformation of $d_A$-modes, which, as mentioned earlier, is achieved by interchanging columns in the Bogoliubov matrix.

The MPS is parametrized by a set of matrices at each site, whose matrix at the $M$-th site is a linear map~\cite{Ostlund1995},
\begin{equation}
	|m_{A}\rangle = \sum_{m_{A-1}}\sum_{n_{M}=0,1} A^{n_M}_{m_{A-1},m_A} |m_{A-1}\rangle \otimes |n_{M}\rangle,
	\label{eq:RG}
\end{equation}
where $|m_{A-1}\rangle$ are the Schmidt vectors for the subsystem $A-1$ (defined by excluding the $M$-th mode from the subsystem $A$) and $|n_M\rangle=(a_{M}^\dagger)^{n_M}|0\rangle_{a_{M}}$ with $n_{M}=0,1$. Since the Schmidt vectors in each subsystem form an orthonormal basis, the MPS matrix for the $M$-th mode
\begin{equation}
	A^{n_M}_{m_{A-1},m_A} =\left( \langle{}n_{M}| \otimes \langle{}m_{A-1}| \right) |m_{A}\rangle
	\label{eq:A_anb}
\end{equation}
takes the form of an overlap between vectors defined in the subsystem $A$.

Before computing the MPS matrix in Eq.~\eqref{eq:A_anb}, we would like to comment on the truncation scheme, as usually the MPS bond dimension (number of Schmidt vectors) would increase exponentially with the system size. From the DMRG point of view, one should keep a manageable number of Schmidt vectors maximizing bipartite entanglement entropy (to minimize the truncation error). From Eq.~\eqref{eq:rhoA}, it is transparent that a Bogoliubov mode with $\Lambda_{A,p}$ close to $0$ or $1$ has less contributions to the entanglement entropy, because the associated two-level system is almost in a pure state (i.e., empty or occupied). Thus, one could keep the Schmidt vectors with the modes being in an empty (occupied) state if $\Lambda_{A,p} < \epsilon$ ($\Lambda_{A,p} > 1-\epsilon$), where $\epsilon$ is the truncation threshold. This ``mode'' truncation scheme preserves the Gaussian nature of the wave function and also arose in a related context~\cite{Fishman2015,Schuch2019}. In our benchmark examples below, we shall adopt this truncation scheme. Alternatively, one could follow the spirit of DMRG and remove the Schmidt vectors whose corresponding eigenvalues of $\rho_A$ are below a certain threshold.

Now we proceed to derive an explicit form for the MPS matrix in Eq.~\eqref{eq:A_anb}. The main difficulty stems from the fact that $|m_{A-1}\rangle$ and $|m_{A}\rangle$ are built upon different Bogoliubov vacua $|0\rangle_{d_{A-1}}$ and $|0\rangle_{d_A}$, respectively. To provide a unified framework, we shall adopt the aforementioned point that both $|m_{A-1}\rangle$ and $|m_{A}\rangle$ can be viewed as Bogoliubov vacua determined by updated Bogoliubov matrices that are obtained by a series of column interchanges on those for $|0\rangle_{d_{A-1}}$ and $|0\rangle_{d_A}$. Without loss of generality, we illustrate below how to calculate Eq.~\eqref{eq:A_anb} for $|m_{A-1}\rangle = |0\rangle_{d_{A-1}}$ and $|m_{A}\rangle = |0\rangle_{d_{A}}$.

With the help of the Bloch-Messiah decomposition~\cite{bloch-messiah}, the Bogoliubov vacua can be rewritten in a form that is more convenient for computing overlaps~\cite{Bertsch2012,Carlsson2021}. As an example, consider $|0\rangle_{d_A}$ that is annihilated by all $d_{A,p}$, $U_A$ and $V_A$ in the Bogoliubov matrix can be decomposed as $U_A=D_A\bar{U}_AC_A$ and $V_A=D_A^*\bar{V_A}C_A$~\cite{bloch-messiah}, where $D_A$ and $C_A$ are unitary matrices. $\bar{U}_A$ and $\bar{V}_A$ are given by
\begin{equation}
	\bar{U}_A =
	\begin{pmatrix}
		\mathds{1} & & \\
		& \bigoplus_p  u_p \sigma^0   & \\
		& & \mathbf{0}  \\
	\end{pmatrix}, \quad
	\bar{V}_A =
	\begin{pmatrix}
		\mathbf{0} & & \\
		& \bigoplus_{p} i v_p \sigma^y  & \\
		& & \mathds{1} \\
	\end{pmatrix},
\end{equation}
where $\mathds{1}$ and $\mathbf{0}$ represent identity and null blocks, $\sigma^0$ and $\sigma^y$ denote the $2\times2$ identity and Pauli matrices, and $u_p$ and $v_p$ are positive numbers satisfying $u_p^2+v_p^2 = 1$. Such a Bloch-Messiah decomposition can be viewed as three successive canonical transformations
\begin{equation}
	\begin{pmatrix}
		d^{\dagger}_A & d_A
	\end{pmatrix}
	=
	\begin{pmatrix}
		a^{\dagger } & a
	\end{pmatrix}
	\begin{pmatrix}
		D_A & 0 \\
		0 & D^{\ast}_A
	\end{pmatrix}
	\begin{pmatrix}
		\bar{U}_A & \bar{V}_A \\
		\bar{V}_A & \bar{U}_A
	\end{pmatrix}
	\begin{pmatrix}
		C_A & 0 \\
		0 & C^{\ast}_A
	\end{pmatrix}.
\end{equation}
By defining the $b$-modes as $b^\dag = a^\dag D_A$, the identity block $\mathds{1}$ (null block $\textbf{0}$) in $\bar{V}_A$ corresponds to fully occupied (empty) $b$-modes in the Bogoliubov vacuum $|0\rangle_{d_A}$, while the remaining $b$-modes appear in a paired form. Given that $b$- and $a$-modes share the same vacuum $|0\rangle_b = |0\rangle_a$, the Bogoliubov vacuum $|0\rangle_{d_A}$ can be rewritten as~\cite{ring-schuck-book}
\begin{equation}
	|0\rangle_{d_A} =\prod_{k\in{}\text{O}}b^\dagger_k\prod_{p\in{}\text{P}}(u_p+v_p{}b^\dagger_{p}b^\dagger_{-p})|0\rangle_a \,,
	\label{eq:vacua_b}
\end{equation}
where $\text{O}$ ($\text{P}$) denotes the set of fully occupied (paired) $b$-modes. By introducing the $f$-modes as $f_p=-v_p b^\dagger_{-p}+u_p{}b_p$ and $f_{-p}=v_p b^\dagger_{p}+u_p{}b_{-p}$ for $p\in\text{P}$ and $f_k=b^\dagger_k$ for $k\in\text{O}$, Eq.~\eqref{eq:vacua_b} can be further rewritten as
\begin{equation}
	|0\rangle_{d_A} = \frac{1}{\prod_{p\in{}\text{P}}v_p}\prod_{k\in\text{O}}f_k\prod_{p\in\text{P}}f_pf_{-p}|0\rangle_a \,,
	\label{eq:vacua_f}
\end{equation}
where $f=a^{\dagger}D_A\bar{V}_A+aD^{\ast}_A\bar{U}_A$. Certainly, one could derive a similar form for $|0\rangle_{d_{A-1}}$ with some $f^{\prime}$-modes acting on the vacuum of $a$-modes in the subsystem $A-1$. With the graded Hilbert space of fermions in mind, $|0\rangle_{d_{A-1}} \otimes |n_{M}\rangle$ is expressed as
\begin{equation}
	|0\rangle_{d_{A-1}} \otimes |n_{M}\rangle = \frac{1}{\prod_{p\in{}\text{P}^{\prime}}v^{\prime}_p}\prod_{k\in\text{O}^{\prime}}f^{\prime}_k\prod_{p\in\text{P}^{\prime}}f_p^{\prime} f_{-p}^{\prime} (a^\dag_{M})^{n_M}|0\rangle_a
	\label{eq:vacua_fa}
\end{equation}
with $f^{\prime}=a^{\dagger}D^{\prime}_A\bar{V}^{\prime}_A+aD^{\ast\prime}_A\bar{U}^{\prime}_A$~\footnote{For convenience, $a^\dag_{M}$ and $a_{M}$ have been included in the definition of $f^{\prime}=a^{\dagger}D^{\prime}_A\bar{V}^{\prime}_A+aD^{\ast\prime}_A\bar{U}^{\prime}_A$. Hence, for $n_M = 1$, $a^{\dagger}_M$ in Eq.~\eqref{eq:vacua_fa} can be viewed as an ``unpaired'' $f^{\prime}$-mode. The Bloch-Messiah decomposition of the Bogoliubov matrix defining $|0\rangle_{d_{A-1}}$ leads to $U_{A-1}=D_{A-1}\bar{U}_{A-1}C_{A-1}$ and $V_{A-1}=D^{\ast}_{A-1}\bar{V}_{A-1}C_{A-1}$. Thus, for those $f^{\prime}$-modes coming from $|0\rangle_{d_{A-1}}$, we have $\bar{V}^{\prime}_A = \bar{V}_{A-1}$, $\bar{U}^{\prime}_A = \bar{U}_{A-1}$, and $D^{\prime}_A$ is obtained from $D_{A-1}$ by appending a trivial row of zeros.}.
By using Wick's theorem, the overlap between $\langle{}n_{M}| \otimes {_{d_{A-1}}}\langle{}0|$ and $|0\rangle_{d_A}$ can be calculated with the following Pfaffian formula~\cite{Bertsch2012}:
\begin{equation}
	\langle f_1^{\prime\dagger} \cdots f_{\mathcal{K}^{\prime}}^{\prime\dagger} f_1 \cdots f_{\mathcal{K}} \rangle_{a}
	= \mathrm{Pf}
	\begin{pmatrix}
		\bar{V}^{\prime T}_A \bar{U}^{\prime}_A & \bar{V}^{\prime T}_A D^{\prime\dagger}_A D_A \bar{V}_A \\
		-\bar{V}^T_A D^T_A D^{\prime\ast}_A \bar{V}^{\prime}_A & \bar{U}^T_A \bar{V}_A
	\end{pmatrix},
	\label{eq:pfaffian}
\end{equation}
where the expectation value is evaluated with respect to $|0\rangle_{a}$, and $f$ and $f^{\prime}$-modes are those appearing in Eqs.~\eqref{eq:vacua_f} and \eqref{eq:vacua_fa} (rows and columns of the matrix within the Pfaffian are selected in accordance with them). It is obvious that Eq.~\eqref{eq:pfaffian} results in a nonzero value only if $\mathcal{K}$ and $\mathcal{K}^{\prime}$ have the same parity, otherwise the overlap vanishes.

Regarding the Pfaffian method, several further comments are in order: (i) The fermion parity symmetry is automatically encoded in each MPS matrix, which is due to the fact that the overlap between $\langle{}n_{M}| \otimes \langle{}m_{A-1}|$ and $|m_{A}\rangle$ vanishes if they have different fermion parities. (ii) The calculation of the MPS matrix [Eqs.~\eqref{eq:A_anb} and \eqref{eq:pfaffian}] can be parallelized, in analogy with the case of Slater determinants~\cite{Petrica2021}. (iii) The (uncontrolled) phase of each Schmidt vector does not play a role. According to Eq.~\eqref{eq:A_anb}, the Schmidt vector $|m_{A}\rangle$ used as a ket for generating $A^{n_M}_{m_{A-1},m_A}$ will be reused as a bra for calculating $A^{n_{M+1}}_{m_{A},m_{A+1}}$, so the phase cancels.

\begin{figure}[tb!]
	\centering
	\includegraphics[width=0.95\linewidth]{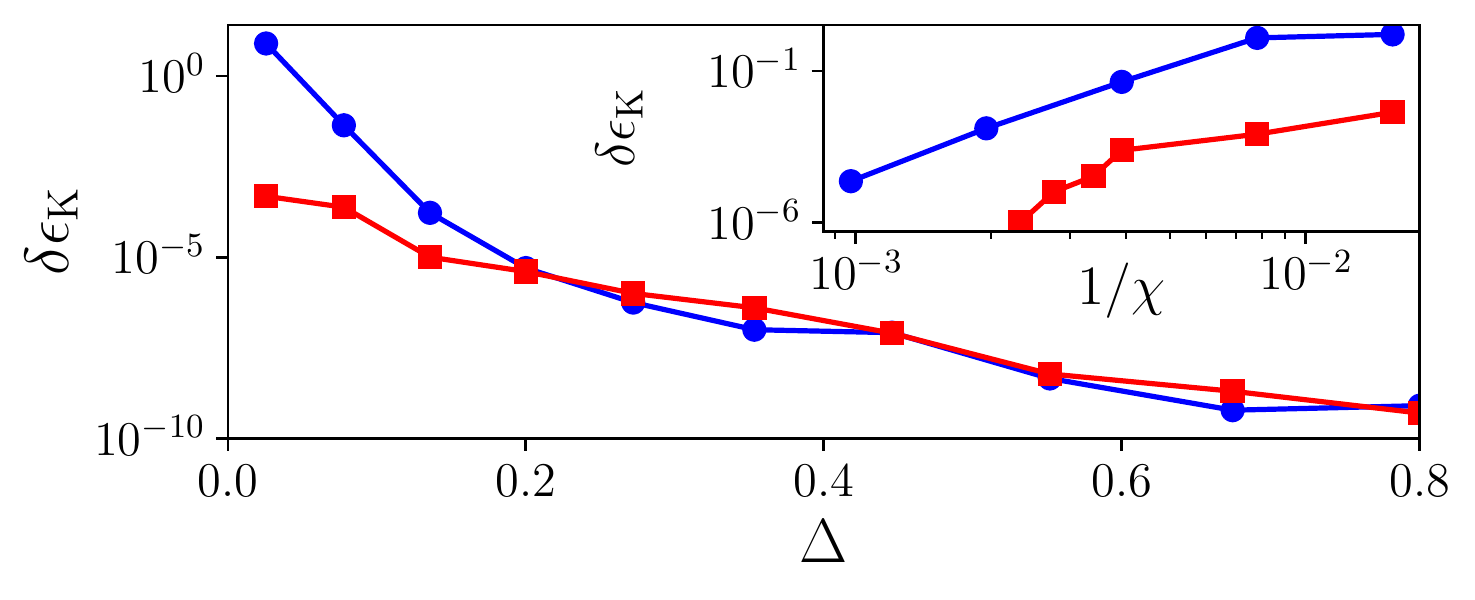}
	\caption{The per-site energy deviation $\delta\epsilon_{\rm{K}}$ as a function of the pairing strength $\Delta$ of the Kitaev chain~\eqref{eq:KitaevChain} obtained by the Pfaffian (red square) and MPO-MPS (blue dot) methods. The Kitaev chain has length $N=60$ with $\mu=0.5$. The same bond dimension $\chi$ is chosen for both methods ($\chi=256$ for $\Delta<0.2$ and $\chi=64$ for $\Delta \geq 0.2$).
		Inset: $\delta\epsilon_{\rm{K}}$ versus $1/\chi$ for $\Delta=0.025$ and $\mu=0.5$.}
	\label{fig:dek}
\end{figure}

{\em Example 1} ---
The Kitaev chain~\cite{Kitaev2001}, whose ground state is a quintessential HFB wave function, is an excellent example to benchmark the Pfaffian method.
The Hamiltonian of the Kitaev chain is parameterized by hopping integral $t$, pairing strength $\Delta$, and chemical potential $\mu$ as follows:
\begin{equation}
	\mathcal{H}_{\rm{K}}=\sum_{j=1}^{N}\left(t{}a_{j}^{\dagger}a_{j+1}+\Delta{}a_{j}a_{j+1}+\mathrm{h.c}\right)+\sum_{j=1}^{N}\mu{}a_{j}^{\dagger}a_{j} \,.
	\label{eq:KitaevChain}
\end{equation}
For simplicity, we set $t=1$ and choose $N = 60$ with antiperiodic boundary condition. When $\Delta=0$, the ground state of \eqref{eq:KitaevChain} is a gapless Fermi sea, whereas a finite $\Delta$ opens an energy gap and turns the system into a topological superconductor for $|\mu|< 2t$.

The ground state of $\mathcal{H}_{\rm{K}}$ defined in Eq.~\eqref{eq:KitaevChain} has been converted to an MPS by using the Pfaffian method as well as the MPO-MPS method~\cite{Jin2020}. The per-site energy deviation $\delta\epsilon_{\rm{K}}$, which is the difference between the value computed from the resulting MPSs and the exact value, is shown in Fig.~\ref{fig:dek}. For a relatively large pairing strength $\Delta$, both the Pfaffian and MPO-MPS methods work very well and give rise to rather precise results with $\delta\epsilon_{\rm{K}}<10^{-5}$. As $\Delta$ is smaller and smaller, the gap of the Kitaev chain is closing, the MPO-MPS method will become less and less accurate, whereas the Pfaffian method still keeps good performance. For both methods, the energy deviations decrease in a power law of the inverse bond dimension $1/\chi$ as $\delta\epsilon_{\rm{K}}\propto{}(1/\chi)^{\alpha}$ (with $\alpha>0$), but it turns out that the Pfaffian method leads to a much larger $\alpha$, i.e., a much more steep descent curve of $\delta\epsilon_{\rm{K}}$ versus $1/\chi$ (see the inset of Fig.~\ref{fig:dek}), suggesting the superiority of the Pfaffian method.

\begin{figure}[tb!]
	\centering
	\includegraphics[width=8.2cm]{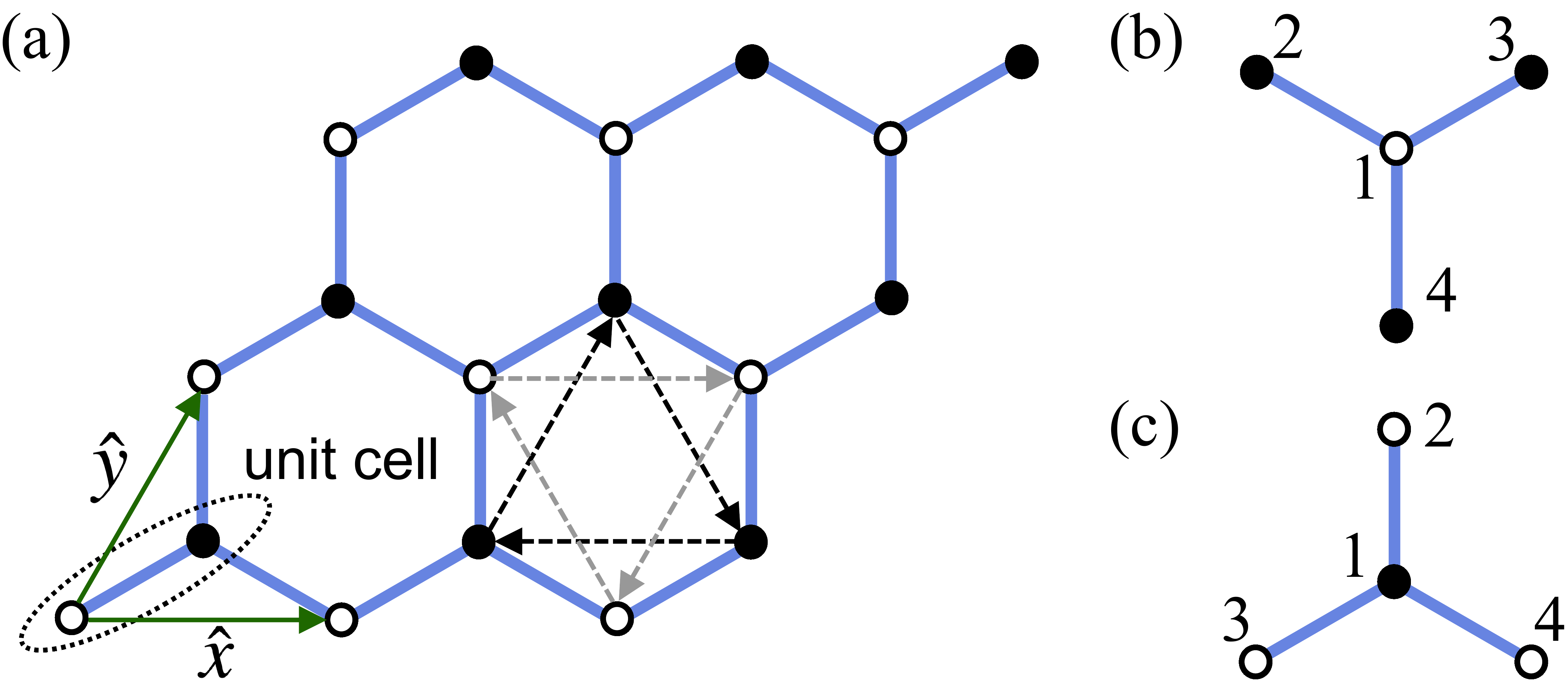}
	\caption{(a) The honeycomb lattice and its unit vectors $\hat{x}$ and $\hat{y}$. The circles (dots) denote the $A$ ($B$) sublattice. The arrows denotes the sign convention in the mean-field Hamiltonian~\eqref{eq:HMF}. (b), (c) The ordering convention of the Majorana operators in (b) $\Ydown$-type and (c) $\Yup$-type four-site Hubbard interactions [see Eq.~\eqref{eq:MajoranaHubbard}], respectively.}
	\label{fig:honeycomb}
\end{figure}

{\em Example 2} ---
The second benchmark example is the Majorana-Hubbard model on the honeycomb lattice~\cite{Franz2018} with $L_x$ ($L_y$) unit cells along the $\hat{x}$ $(\hat{y})$ directions, as illustrated in Fig.~\ref{fig:honeycomb}(a).
The interactions of Majorana fermions $\alpha_i$ ($\beta_j$) on the $A$ ($B$) sublattice are described by four-Majorana-fermion terms and result in the Hamiltonian
\begin{equation}
	\mathcal{H}_{\text{M}}=i\sum_{\langle{}ij\rangle}\alpha_i\beta_j+g\left(\sum_{ijkl\in\Ydown}\alpha_i\beta_j\beta_k\beta_l+\sum_{{ijkl}\in\Yup}\beta_i\alpha_j\alpha_k\alpha_l\right),
	\label{eq:MajoranaHubbard}
\end{equation}
where $\langle{}ij\rangle$ denotes the nearest-neighbor Majorana hybridization, and $ijkl\in\Ydown$ ($\in\Yup$) denotes the four-site Majorana-Hubbard interactions on the $\Ydown$-type ($\Yup$-type) stars [see Figs.~\ref{fig:honeycomb}(b) and \ref{fig:honeycomb}(c) for the ordering convention].

In the noninteracting limit ($g=0$), the model~\eqref{eq:MajoranaHubbard} is exactly solvable and has a gapless Dirac point in the first Brillouin zone. We introduce a spinless complex fermion in each unit cell $\bf{r}$, $a^\dag_{\bf{r}}=(\alpha_{\bf{r}}-i\beta_{\bf{r}})/2$ and $a_{\bf{r}}=(\alpha_{\bf{r}}+i\beta_{\bf{r}})/2$, and then construct the ground state in the basis of $a_{\bf{r}}$-fermions by using our Pfaffian method. Numerical computations have been carried out on an $L_x\times{}L_y=20\times6$ lattice with MPS bond dimension $\chi$ up to 1024. The cylindrical boundary condition is adopted with $\hat{y}$-direction being periodic. For $g=0$, the energy deviation of the resulting MPS is about $5\times{}10^{-5}$, indicating that the Pfaffian method still works quite well.

The renormalization group analysis in Ref.~\cite{Franz2018} pointed out that a weak interaction $g$ is sufficient to open a gap for the Majorana-Hubbard model~\eqref{eq:MajoranaHubbard} and results in a topological superconductor with spectral Chern number $-\text{sgn}(g)$. It was suggested that the low-energy physics of the Hamiltonian~\eqref{eq:MajoranaHubbard} is captured by the mean-field Hamiltonian~\cite{Franz2018}
\begin{equation}
	H_{M}=i\sum_{\langle{}ij\rangle}\alpha_i\beta_j+it'\sum_{\langle\langle{}kl\rangle\rangle}\eta_{kl}\left(\alpha_k\alpha_l+\beta_k\beta_l\right),
	\label{eq:HMF}
\end{equation}
where $\langle\langle{}kl\rangle\rangle$ denotes next-nearest-neighbor (NNN) bonds and $\eta_{kl}=\pm{}1$ is the sign structure for the NNN hoppings [see Fig.~\ref{fig:honeycomb}(a)]. With a nonzero $t'$, the system is gapped, belongs to class D in the free-fermion classification~\cite{Altland1997,Kitaev2009,Ryu2010,Chiu2016}, and has a spectral Chern number $\text{sgn}(t')$. Therefore, the ground state of the Hamiltonian~\eqref{eq:HMF} is a HFB wave function analogues to the $p\pm{}ip$ topological superconductor.

\begin{figure}
	\centering
	\includegraphics[width=0.95\linewidth]{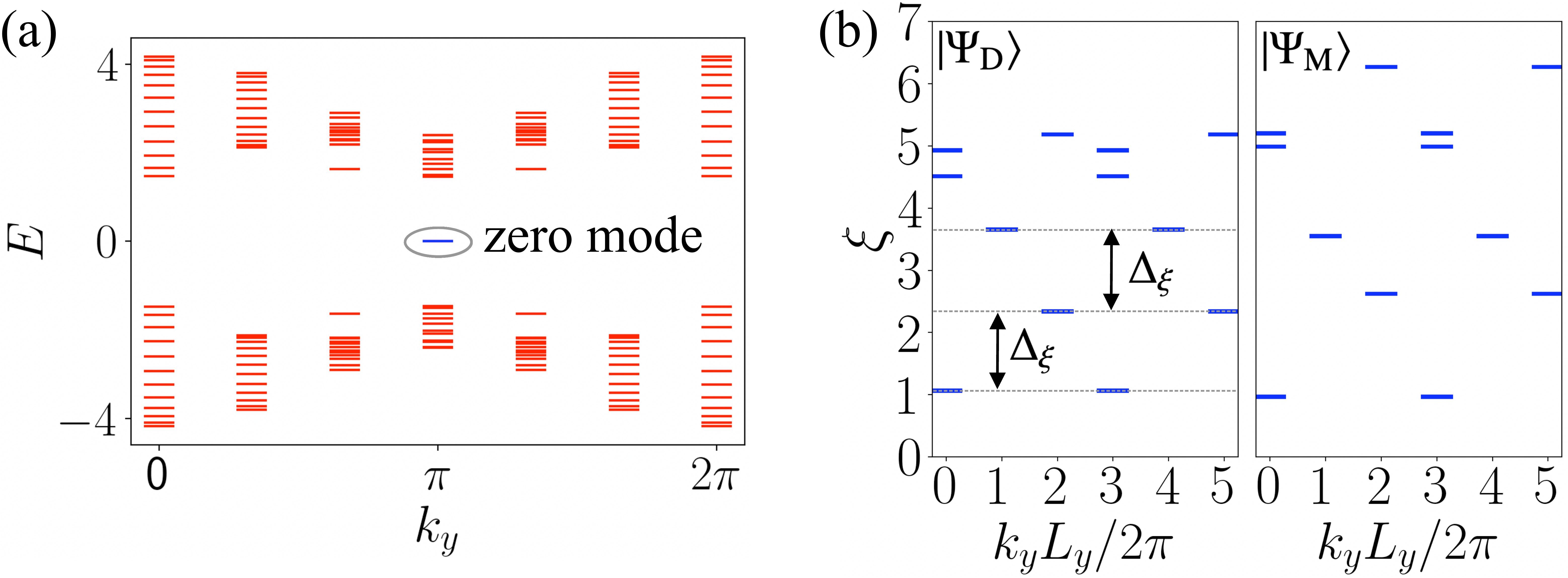}
	\caption{(a) The single-particle spectrum of the Hamiltonian~\eqref{eq:HMF} on a cylinder with $t'=-0.2$, $L_x= 12$, and $L_y= 6$. There is one complex fermion zero mode at $k_y=\pi$. (b) The entanglement spectrum of $|\Psi_{\rm{M}}\rangle$ and $|\Psi_{\rm{D}}\rangle$ versus $k_y$. The dashed lines indicate the equidistant entanglement energy levels observed in $|\Psi_{\rm{D}}\rangle$.}
	\label{fig:ee}
\end{figure}

The Hamiltonian~\eqref{eq:HMF} on a cylinder with $L_x$ being large hosts a complex fermion zero mode [see Fig.~\ref{fig:ee}(a)], which composes of two Majorana modes at the two boundaries of the cylinder.
Leaving this mode empty or occupied gives rise to two topologically degenerate ground states with different fermion parity. For either state, the entanglement spectrum (i.e., the negative logarithm of the eigenvalues of reduced density matrix~\cite{Li2008}) exhibits (at least) two-fold degeneracy~\cite{Stephan2009,Fidkowski2010,Dubail2011}. This nontrivial topological feature is captured by the Pfaffian method, since it precisely preserves the degeneracy in the reduced density matrix [see Eq.~\eqref{eq:rhoA}]. We determine the ground state $|\Psi_{\rm M}\rangle$ of the mean-field Hamiltonian~\eqref{eq:HMF} in the even-fermion-parity sector on an $L_x\times L_y=12\times 6$ cylinder, and use the Pfaffian method to approximate it as an MPS with bond dimension $\chi_{\rm M}=600$. Here we have chosen $t'=-0.2$ in Eq.~\eqref{eq:HMF}, which corresponds to the mean-field solution to the Majorana-Hubbard model~\eqref{eq:MajoranaHubbard} at $g=0.5$~\cite{Franz2018}. Figure~\ref{fig:ee}(b) clearly shows the doubly degenerate entanglement spectrum of $|\Psi_{\rm M}\rangle$ at each entanglement energy level.

By combining the Pfaffian and DMRG methods, we are able to find out the actual ground state of the Majorana-Hubbard model~\eqref{eq:MajoranaHubbard} on the same $12\times{}6$ cylinder with coupling strength $g=0.5$ and simultaneously diagnose the quality of $|\Psi_{\rm M}\rangle$ as its variational ansatz. The ground state $|\Psi_{\rm D}\rangle$ of the Hamiltonian~\eqref{eq:MajoranaHubbard} is obtained by using the DMRG method with $|\Psi_{\rm M}\rangle$ being the initial ansatz, and the bond dimension for DMRG calculations is $\chi_{\rm D}=3000$.
The relative energy difference between $|\Psi_{\rm M}\rangle$ and $|\Psi_{\rm D}\rangle$ is somewhat large (${\sim}\,6.5\%$) and the fidelity reads $|\langle\Psi_{\rm D}|\Psi_{\rm M}\rangle|\approx{}0.499$. Similar to $|\Psi_{\rm M}\rangle$, the DMRG-optimized state $|\Psi_{\rm D}\rangle$ still sustains a doubly degenerate entanglement spectrum, as shown in Fig.~\ref{fig:ee}(b). Moreover, the counting $\{1, 1, 1, 2, \dots\}$ of one chiral branch in the entanglement spectrum of $|\Psi_{\rm D}\rangle$ is consistent with the Ramond sector of the free Majorana fermion conformal field theory, indicating that $|\Psi_{\rm D}\rangle$ supports a chiral Majorana edge mode at each boundary and the Hamiltonian~\eqref{eq:MajoranaHubbard} hosts an interacting topological superconductor at the relatively large coupling strength $g=0.5$.

For the same model, we also carry out the DMRG ground-state search with random MPSs as initial ansatz and obtain a converged MPS $|\Psi_{\rm R}\rangle$. Unlike that of $|\Psi_{\rm D}\rangle$, the entanglement spectrum of $|\Psi_{\rm R}\rangle$ is no longer exactly doubly degenerate, where the relative difference between two lowest entanglement energy levels is ${\sim}\,10^{-4}$. The variational energy of $|\Psi_{\rm R}\rangle$ is also slightly higher than that of $|\Psi_{\rm D}\rangle$ with a relative difference ${\sim}\,10^{-4}$. This comparative study implies that although the HFB state $|\Psi_{\rm M}\rangle$ quantitatively differs from the actual many-body ground state of the Hamiltonian~\eqref{eq:MajoranaHubbard}, it nevertheless captures the essential physics. Even if $|\Psi_{\rm M}\rangle$ comes from an empirical mean-field Hamiltonian and is not fully optimized in the sense of the HFB theory, this example has already demonstrated that combining HFB and DMRG methods is promising to approach quantum many-particle systems.

{\em Summary and outlook} --- To summarize, we have put forward the Pfaffian method for converting HFB wave functions into MPSs. This approach is generally more accurate than previous methods and is easily parallelizable, as demonstrated by our numerical studies on the Kitaev chain as well as the Majorana-Hubbard model on the honeycomb lattice. To give some perspective, the Pfaffian method could serve as a hub between the HFB and DMRG methods, as the correlation matrix, which is the optimized output of the HFB method (see, e.g., Ref.~\cite{Kraus2010}), can be directly used for producing an MPS and subsequently be supplied to DMRG for further improvements. This would take advantage of both methods and avoid certain shortcomings of each. Given the wide applications of HFB and DMRG methods, the combination via the Pfaffian method is a promising direction for future investigations.

\begin{acknowledgments}
	{\em Acknowledgments} --- We are grateful to Jan von Delft, Xiao-Liang Qi, Lei Wang, Ying-Hai Wu, and Qi Yang for stimulating discussions. H.-K.J. is funded by the European Research Council (ERC) under the European Unions Horizon 2020 research and innovation program (Grant Agreement No. 771537). Y.Z. is supported by National Natural Science Foundation of China (No. 12034004 and No. 11774306) and the Strategic Priority Research Program of Chinese Academy of Sciences (No. XDB28000000). H.-H.T. is supported by the Deutsche Forschungsgemeinschaft (DFG) through project A06 of SFB 1143 (project-id 247310070). The numerical simulations in this work are based on the GraceQ project~\cite{graceq} and TeNPy Library~\cite{tenpy}.
\end{acknowledgments}

\bibliography{HFB}

\end{document}